\documentclass[reqno, 12 pt]{amsart}
\usepackage{amsmath,amssymb,amsthm,
	amsfonts,latexsym}
\usepackage{hyperref}
\usepackage{txfonts}
\usepackage{color}

\numberwithin{equation}{section}

\theoremstyle{plain}

\theoremstyle{definition}

\theoremstyle{remark}

\newcounter{dc}

\DeclareMathOperator{\E}{e}

\def\ulamek#1#2{\mbox{\normalfont$\frac{#1}{#2}$}}

\begin{document}

\title[Lacunary Generating Functions for the Laguerre Polynomials]
{Lacunary Generating Functions for the Laguerre Polynomials}

\author{D. Babusci}
\address{INFN - Laboratori Nazionali di Frascati, via E. Fermi, 40, IT 00044 Frascati (Roma), Italy}
\email{danilo.babusci@lnf.infn.it}

\author{G. Dattoli}
\address{ENEA - Centro Ricerche Frascati, via E. Fermi, 45, IT 00044 Frascati (Roma), Italy}
\email{dattoli@frascati.enea.it}

\author{K. G\'{o}rska}
\address{H. Niewodnicza\'{n}ski Institute of Nuclear Physics, Polish Academy of Sciences, ul.\ Eliasza-Radzikowskiego 152, PL 31-342 Krak\'{o}w, Poland}
\email{katarzyna.gorska@ifj.edu.pl}

\author{K. A. Penson}
\address{Laboratoire de Physique Th\'{e}orique de la Mati\`{e}re Condens\'{e}e, Sorbonne Universit\'{e}s, Universit\'{e} Pierre et Marie Curie, CNRS UMR 7600 Tour 13 - 5i\`{e}me \'{e}t., B.C. 121, 4 pl.\ Jussieu, F 75252 Paris Cedex 05, France}
\email{penson@lptl.jussieu.fr}



\begin{abstract}
Symbolic methods of umbral nature play an important and increasing role in the theory of special functions and in related fields like combinatorics. We discuss an application of these methods to the theory of lacunary generating functions for the Laguerre polynomials for which we give a number of new closed form expressions. We present furthermore the different possibilities offered by the method we have developed, with particular emphasis on their link to a new family of special functions and with previous formulations, associated with the theory of quasi monomials.
\end{abstract}

\maketitle

\thispagestyle{myheadings}
\font\rms=cmr8 
\font\its=cmti8 
\font\bfs=cmbx8

\def\thepage{}

\section{Introduction}

\subsection{Motivation}

This work deals with derivation of a number of summation formulas involving ordinary and generalized Laguerre polynomials of degree 
$n$, denoted by $L_{n}(x)$ and $L_{n}^{(\alpha)}(x)$, respectively. More specifically, we are interested in ordinary, exponential and more general generating functions for polynomials 
$L_{2n}(x)$, $L_{3n}(x)$, $L_{2n+l}(x)$, $L_{3n+l}(x)$ etc., and for their generalized counterparts. In this paper they will be called 
\emph{lacunary} generating functions. 
The Laguerre polynomials appear quite naturally in the theory of the following differential operator
\begin{equation}\label{30/10/2014-1}
	D^{(\alpha)}_{x} = \frac{d}{d x} x \frac{d}{d x} + \alpha \frac{d}{d x},
\end{equation}
where $x$ is the indeterminate and $\alpha$ is a real parameter
\cite{DCocolicchio99, GDattoli05a, GDattoli04, KAPenson09}. The specific context is a less-known \cite{JDL} formulation of theory of polynomials $P_{n}(x, \lambda)$ (of degree $n$ in $x$) 
which is based on the action of the
exponential of a certain differential operator 
$\hat{O}(x, \ulamek{d}{dx})$ on monomials $x^{n}$ 
through 
\begin{equation}\label{30/10/2014-2}
	{\exp\left[\lambda\, \hat{O}(x, \ulamek{d}{dx})\right]} \, x^{n} = P_{n}(x, \lambda), \quad (n=0, 1, \ldots) .
\end{equation}
Eq.~\eqref{30/10/2014-2} is the 
defining relation of a family of two-variable polynomials and $P_{n}(x, \lambda)$  
appears to be related to  
umbral calculus \cite{Roman, Rota}. For the 
two-variable Hermite polynomials $H^{(2)}_{n}(x, \lambda)$ the following relation holds \cite{DCocolicchio99, GDattoli04}:
\begin{equation}\label{30/10/2014-3}
	{\exp\left[\lambda\, \ulamek{d^{2}}{dx^{2}}\right]}\, x^{n} = H^{(2)}_{n}(x, \lambda),
\end{equation}
where the polynomials
$H^{(2)}_{n}(x, \lambda)$ are related to the conventional Hermite polynomials $H_{n}(x)$ through
\begin{align}\label{30/10/2014-4}
	H^{(2)}_{n}(x, \lambda) &= (-i\sqrt{\lambda})^{n}\, H_{n}(\ulamek{i\, x}{2\sqrt{\lambda}}), & H_{n}(x) &= 2^{n} \,H^{(2)}_{n}(x, -{1/4}).
\end{align}
 
Relation \eqref{30/10/2014-2} for the operator 
$\hat{O}(x, \ulamek{d}{dx}) = D^{(\alpha)}_{x}$ reads (see, e.g., \cite{KAPenson09})
\begin{equation}\label{30/10/2014-5}
	{\exp\left[\lambda\, D^{(\alpha)}_{x}\right]}\, x^{n} = n!\, L^{(\alpha)}_{n}(\lambda, x) = n! \, \lambda^{n} \,L^{(\alpha)}_{n}(-x/{\lambda}),
\end{equation}
which justifies the name `Laguerre derivative' for $D^{(\alpha)}_{x}$ \footnote{Strictly speaking, Laguerre polynomials are one-variable polynomials and are defined as $L_{n}(x) = L_{n}({1, -x})$. It is also easily checked that {$L_{n}(x, y) = x^n L_{n}(-\ulamek{y}{x})$}. We will use two-variable forms for future convenience. In final results we shall again mostly use the one-variable form.}. Eq.~\eqref{30/10/2014-5} was obtained using the fact, easily shown by induction,
\begin{equation}\label{31/10/2014-1}
	\left[D^{(\alpha)}_{x}\right]^{k} x^{n} = \frac{n!}{(n-k)!}\, (1+\alpha-n-k)_k\, x^{n-k},
\end{equation}
with $(a)_{n} = \Gamma(a+n)/\Gamma(a)$ being the Pochhammer symbol. It generalizes Eq.~(22) of \cite{KAPenson09} to $\alpha\neq 0$. As an immediate consequence of Eq.~\eqref{31/10/2014-1} the following relation obtains:
\begin{equation}\label{31/10/2014-2}
	{\exp\left[\lambda \,D^{(\alpha)}_{x}\right]} \E^{-x} = \frac{1}{(1+\lambda)^{1+\alpha}}\exp\left(\frac{-x}{1+\lambda}\right),
\end{equation}
which is equivalent to one of the standard generating functions for the Laguerre polynomials.

In general, the differential operator $\hat{O}(x, \ulamek{d}{dx})$ can be seen as part of the evolution equation for a function $f(x, t)$, given  by a partial differential equation of the following type:
\begin{equation}\label{31/10/2014-3}
	\frac{\partial}{\partial t} \,f(x, t) = \kappa \,\hat{O}(x, \ulamek{\partial}{\partial x})\, f(x, t), \quad x > 0,
\end{equation}
where $\kappa$ is a 
coupling constant (assuming here $\kappa=1$), which for the operator
$\hat{O}(x, \ulamek{d}{dx})\equiv -D^{(\alpha)}_{x}$ becomes the Cauchy problem with the initial condition $f(x)$ \cite{DHR}:
\begin{align}\label{31/10/2014-4}
	\frac{\partial}{\partial t} \,f_{\alpha}(x, t) &= -\left(\frac{\partial}{\partial x} x\frac{\partial}{\partial x} + \alpha \frac{\partial}{\partial x}\right) \,f_{\alpha}(x, t),   \\
\label{31/10/2014-5}
	f_{\alpha}(x, 0) &= f(x).
\end{align}
The formal solution of Eqs.~\eqref{31/10/2014-3} and \eqref{31/10/2014-5} is obtained via 
\begin{equation*}
	f_{\alpha}(x, t) = {\exp\left[-t\,D^{(\alpha)}_{x}\right]}\, f(x).
\end{equation*}
Therefore Eq.~\eqref{31/10/2014-2} for $\lambda \equiv t$ describes the exact time evolution under the Laguerre derivative $D^{(\alpha)}_{x}$ from the initial condition $f(x) = \E^{-x}$. 
Whereas Eq.~\eqref{31/10/2014-2} could have been obtained via conventional and well-known formula for $\sum_{n=0}^{\infty} x^{n} L^{(\alpha)}_{n}(x)$, this is not anymore the case for initial conditions differing from the exponential. 

\pagenumbering{arabic}
\addtocounter{page}{1}
\markboth{\SMALL D. BABUSCI, G. DATTOLI, K. G\'{O}RSKA and K. A. PENSON}{\SMALL LACUNARY GENERATING FUNCTIONS FOR THE LAGUERRE POLYNOMIALS}

We illustrate this situation by choosing, e.g., $f(x) = x^{-\alpha/2} \,I_{\alpha/2}(\beta\, x)$, with $\alpha, \beta~>~0$, respectively $f(x) = I_{0}(\beta\, x)$, where $I_{\alpha/2}(z)$ is the modified Bessel function of the first kind,
\begin{equation*}
	I_{\alpha/2}(\beta \,x) = \sum_{n=0}^{\infty}\frac{ (\beta \, x/2)^{2n+\alpha/2}}{n!\, \Gamma(1+\alpha/2+n)}.
\end{equation*}
We obtain
\begin{align}\label{2/03/2015-a}
	\exp\left[-t\,D^{(\alpha)}_{x}\right]x^{-\alpha/2} I_{\alpha/2}(\beta\, x) &= \frac{(\beta/2)^{\alpha/2}}{\Gamma(1+\alpha/2)} \sum_{n=0}^{\infty} \frac{(1/2)_{n}}{(1+\alpha/2)_n}\, (\beta\, t)^{2n} L^{(\alpha)}_{2 n}(x/t), \\
\label{31/10/2014-7}
	\exp\left[-t\,D^{(\alpha)}_{x}\right]\, I_{0}(\beta\, x) &= \sum_{n=0}^{\infty} \frac{(1/2)_{n}}{n!}\, (\beta\,t)^{2n} \,L^{(\alpha)}_{2 n}(x/t).
\end{align}
Any further evaluation of 
Eqs.~\eqref{2/03/2015-a} and \eqref{31/10/2014-7} turned out to be impossible. In addition, an extensive search in the literature for the appropriate formula gave no results.

We have considered it therefore as our objective to establish the summation formulas of type Eqs.~\eqref{2/03/2015-a} and \eqref{31/10/2014-7}, and more general ones, which involved lacunary Laguerre series. 
(Some lacunary exponential generating functions for other polynomials are known: compare \cite{APPrudnikov98} for $H_{2n}(x)$, \cite{IMGessel05} for $H_{3n}(x)$, and various lacunary generating functions for Legendre and Chebyshev polynomials \cite{LacPeCh}).

Note that the differential equations of type \eqref{31/10/2014-3} were already considered in a 
combinatorial context as a tool to derive generating functions for combinatorial graphs associated with $\hat{O}(x, \ulamek{d}{d x})$, see \cite{KAPenson09}.

\subsection{Description of the method}

It has been shown that symbolic methods of umbral nature provide powerful tools to deal with the properties of special polynomials and functions \cite{DBabusci12b, DBabusci13, DBabusci12a, GDattoli00}. These techniques greatly simplify the problems underlying such studies and allow to reduce the derivation of the relevant properties to straightforward algebraic manipulations.

Similarly to classical references on the umbral calculus \cite{Roman, Rota}, we start our considerations with the shift operator $c_{z}$ defined here by
\begin{equation}\label{11/12/2014-1}
	c^{\alpha}_{z}: f(z) \mapsto f(z+\alpha),
\end{equation}
satisfying $c^{\alpha}_{z} c^{\beta}_{z} = c^{\alpha+\beta}_{z}$. When $c^{\alpha}_{z}$ acts on $[\Gamma(1+z)]^{-1}$, the result evaluated at $z=0$ gives $[\Gamma(1+\alpha)]^{-1}$. In this context the ordinary
Laguerre polynomials $L^{(0)}_{n}(x)~\equiv~L_{n}(x)$ can be obtained with the help of $c^{\alpha}_{z}$ as
\begin{align}\label{11/12/2014-2a}
	L_{n}(x) & = (1 - x c_{z})^{n} \,\frac{1}{\Gamma(1+z)}\Big\vert_{z=0} \\ \nonumber
	& = \sum_{k=0}^{n} \binom{n}{k}\, (-x)^k \,c^{k}_{z} \,\frac{1}{\Gamma(1+z)}\Big\vert_{z=0} \\ \nonumber
	& = \sum_{k=0}^{n} \binom{n}{k} \,(-x)^k \,\frac{1}{\Gamma(1+k+z)}\Big\vert_{z=0} \\ \nonumber
	& = \sum_{k=0}^{n} \binom{n}{k} \,\frac{(-x)^k}{k!}.
\end{align}
In the same vein, the two-variable generalized Laguerre polynomials 
\begin{align}\label{12/12/2014-1}
	L^{(\alpha)}_{n}(x, y)&=\frac{1}{n!}\sum_{k=0}^n\binom{n}{k}\,(1+\alpha+k)_{n-k}\, x^{n-k}\,y^k \cr 
	&= 
	\frac{\Gamma(1+\alpha+n)}{n!}\, \Lambda^{(\alpha)}_{n}(x, y)
\end{align}
are obtained via
\begin{align}
\label{11/12/2014-3a}
	\Lambda^{(\alpha)}_{n}(x, y)&=\sum_{k=0}^{n} \binom{n}{k} \,\frac{x^{n-k}\, y^{k}}{\Gamma(1+\alpha+k)}\\ \nonumber
	& = \sum_{k=0}^{n} \binom{n}{k} \, x^{n-k}\, y^{k} \, c^{k}_{z} \, \frac{1}{\Gamma(1+z)}\Big\vert_{z=\alpha}  \\ \label{11/12/2014-3c}
	&= (x + y\,c_{z})^{n}\, \frac{1}{\Gamma(1+z)}\Big\vert_{z=\alpha} \\ \label{11/12/2014-3aa} 
	& = c_{z}^{\alpha}\,(x+y\, c_{z})^{n} \frac{1}{\Gamma(1+z)}\Big\vert_{z=0}.
\end{align}
Note that with our definition of $L^{(\alpha)}_{n}(x)$ given in Eqs.~\eqref{12/12/2014-1} {with \eqref{11/12/2014-3c} or \eqref{11/12/2014-3aa}} we can do about the same as with the definition \`{a} la Rota (see p.~198 in \cite{Rota})\footnote{We did not attempt to demonstrate rigorously the equivalence of these two definitions.}, namely $L^{(\alpha)}_{n}(x) = x^{-\alpha}\,(\ulamek{d}{d x} - I)^{n} x^{n+\alpha}$.

Using the method exposed in Eqs.~\eqref{11/12/2014-2a} and \eqref{12/12/2014-1}, 
the Laguerre polynomials $L_{2n}(x, y)$ and $L_{3n}(x, y)$ can be expressed via $\Lambda_{n}^{(\alpha)}(x, y)$ as follows: 
\begin{equation}\label{EQ1.4}
	L_{2 n}(x, y) = (x + y \,c_{z})^{2 n} \frac{1}{\Gamma(1+z)}\Big\vert_{z=0} = \sum_{r=0}^{n} \binom{n}{r} \, x^{n-r} \, y^r \,\Lambda_{n}^{(r)}(x, y),
\end{equation}
and
\begin{equation*}
	L_{3 n}(x, y) = (x + y\, c_{z})^{3 n} \frac{1}{\Gamma(1+z)}\Big\vert_{z=0} = \sum_{k=0}^{n} \sum_{r=0}^{n} \binom{n}{r}\, \binom{n}{k} \,x^{2 n-r-k} \, y^{r+k} \,\Lambda_{n}^{(r+k)}(x, y).
\end{equation*}

A further family of polynomials, introduced by the same means as before, is provided by what we will call, for reasons which will become clear in the following, the 
two-parameter family of Laguerre--Wright polynomials, namely
\begin{equation}\label{EQ1.6}
	\Lambda_{n}^{(\alpha, \beta)}(x, y) = (x + y \,c^{\beta}_{z})^n\, \frac{1}{\Gamma(1+z)}\Big\vert_{z=\alpha} = \sum_{r=0}^{n} \binom{n}{r}\, \frac{x^{n-r}\,y^r }{\Gamma(1 + \alpha+\beta\, r)}.
\end{equation}
From Eq.~\eqref{11/12/2014-3c} $\Lambda_{n}^{(\alpha, 1)}(x, y) = \Lambda_{n}^{(\alpha)}(x, y)$ follows.

\section{Standard generating functions of Laguerre polynomials}

The first family of examples will concern standard generating functions of polynomials $\Lambda^{(\alpha)}(x)$ and $\Lambda^{(\alpha,\beta)}(x)$. In this way we illustrate the usefulness of our approach. 
We will consider in the following two types of generating  functions, namely the exponential and the ordinary ones. In the first case, Eq.~(\ref{EQ1.6}) leads to the following form
\begin{align}\label{EQ1.7}
	\sum_{n=0}^{\infty} \frac{t^n}{n!} \,\Lambda_{n}^{(\alpha, \beta)}(x, y) & = \sum_{n=0}^{\infty} \frac{1}{n!} [t \,(x + y \,c^{\beta}_{z})]^{n}\, \frac{1}{\Gamma(1+z)}\Big\vert_{z=\alpha} \nonumber \\
	& = \E^{x t}\, \exp\left[{ t\,y\, c^{\beta}_{z}}\right] \frac{1}{\Gamma(1+z)}\Big\vert_{z=\alpha} \nonumber \\ 
	& = \E^{x t}  W^{(\beta, \alpha+1)}(t\, y),
\end{align}
where the Bessel--Wright function $W^{(\beta, \alpha)}(x)$ \cite{DBabusci11, DBabusci12, IPodlubny99} reads
\begin{align*}
	W^{(\beta, \alpha)}(x) &= \exp\left[x \,c^{\beta}_{z}\right] \frac{1}{\Gamma(1+z)}\Big\vert_{z=\alpha-1} \\ 
	&= \sum_{r=0}^{\infty} \frac{x^r}{r!}\, c_z^{\beta\, r} \frac{1}{\Gamma(1+z)}\Big\vert_{z=\alpha-1} \\
	&= \sum_{r=0}^{\infty} \frac{x^r}{r! \,\Gamma(\alpha+\beta\, r)}.
\end{align*}
Observe that $W^{(\beta, \alpha)}(x)$ for $\beta=1$ is equal to $[\Gamma(\alpha)]^{-1} {_{0}F_{1}}\left({\genfrac {}{}{0pt}{} - \alpha} ; x\right)$. Hence Eq.~\eqref{EQ1.7} can be rewritten as
\begin{align*}
	\sum_{n=0}^{\infty} \frac{t^n}{(1+\alpha)_{n}} L^{(\alpha)}_{n}(1, -y) &= {\E^{t}}\, {_{0}F_{1}}\left({\genfrac {}{}{0pt}{}-  {1+\alpha}}; {-t y}\right) \cr
	&= \Gamma(1+\alpha)\, {(y\,t)^{-\alpha/2}\, \E^{t}\, J_{\alpha}(2\sqrt{y\, t})},
\end{align*}
where $J_{\alpha}(z)$ is the Bessel function, see Eq.~(9.12.11) on p.~242 of \cite{Koekoe} or Eq.~(5.11.2.5) for $n=0$ on p.~704 of \cite{APPrudnikov98}.

In the case of ordinary generating function, Eq.~(\ref{EQ1.6}) gives 
\begin{align}\label{EQ1.9}
	\sum_{n=0}^{\infty} t^n\, \Lambda_{n}^{(\alpha, \beta)}(x, y) & = \sum_{n=0}^{\infty} [t\, (x + y\, c^{\beta}_{z})]^{n} \frac{1}{\Gamma(1+z)}\Big\vert_{z=\alpha} \nonumber\\ 
	& = \left[\frac{1}{1-t\, (x + y\,c^{\beta}_{z})}\right]\, \frac{1}{\Gamma(1+z)}\Big\vert_{z=\alpha} \nonumber \\
	& =  \frac{1}{1-t\, x}\cdot\left[\frac{1}{1-\frac{t\, y}{1-t \,x} c^{\beta}_{z}}\right] \,\frac{1}{\Gamma(1+z)}\Big\vert_{z=\alpha} \nonumber \\ 
	& = \frac{1}{1-t\, x} E_{\beta, \alpha +1} \left(\frac{t\,y}{1-t\,x}\right),
\end{align}
where $E_{\beta, \alpha}(x)$ is the two-parameter Mittag-Leffler function \cite{DBabusci11, DBabusci12, IPodlubny99} defined as
\begin{align}\label{EQ1.10}
	E_{\beta, \alpha}(x) & = \left[\frac{1}{1-x\,c^{\beta}_{z}} \right]\frac{1}{\Gamma(1+z)}\Big\vert_{z=\alpha-1} \nonumber \\
	& = \sum_{r=0}^{\infty} x^r\, c^{\beta \,r}_{z} \frac{1}{\Gamma(1+z)}\Big\vert_{z=\alpha-1} \nonumber \\
	& = \sum_{r=0}^{\infty} \frac{x^r}{\Gamma(\alpha+\beta\, r)},
\end{align}
The function introduced in Eq.~\eqref{EQ1.10} for $\beta = 1$ is equal to $[\Gamma(\alpha)]^{-1}\, {_{1}F_{1}}\left({\genfrac {}{}{0pt}{}1 \alpha}; x \right)$. Thus Eq.~\eqref{EQ1.9} for $\beta=1$ goes over to 
Eq.~(9.12.12) for $\gamma=1$ on p.~242 of \cite{Koekoe}, or Eq.~(5.11.2.6) for $b = 1$ on p.~704 of \cite{APPrudnikov98}, that is,
\begin{equation*}
	\sum_{n=0}^{\infty} \frac{n!}{(1+\alpha)_{n}}\, t^{n}\, {L^{(\alpha)}_{n}(1, y)} = \frac{1}{1 - t}\, {_{1}F_{1}}\left({\genfrac {}{}{0pt}{}1 {1+\alpha}}; {\frac{t \,y}{1 - t}}\right).
\end{equation*}


The ordinary and exponential generating functions of the {generalized} Laguerre polynomials are accordingly derived as the special case of previous demonstrations; defining\,\footnote{The subscripts in $G$'s indicate the degree of lacunarity: 1-no lacunarity, 2-double-lacunary, etc; see also the next paragraph.} for $\beta=1$:
\begin{equation}\nonumber
	{_{O}G}_{1}^{(\alpha, 1)}(x, y \,|\,t) = \sum_{n=0}^{\infty} t^n\, L_{n}^{(\alpha)}(x, y) \quad \text{and} \quad {_{E}G}_{1}^{(\alpha, 1)}(x, y \,|\,t) = \sum_{n=0}^{\infty} \frac{t^n}{n!}\, L_{n}^{(\alpha)}(x, y),
\end{equation}
we obtain from 
(\ref{12/12/2014-1}), (\ref{11/12/2014-3c}) and the binomial series expansion:
\begin{align}\label{EQ1.11}
	{_{O}G}_{1}^{(\alpha, 1)}(x, y \,|\,t) &= \Gamma(1+\alpha)\left[\sum_{n=0}^\infty\frac{(1+\alpha)_n}{n!} (x+y\,c_z)^n\right]\frac{1}{\Gamma(1+z)}\Big\vert_{z=\alpha}\nonumber\\
	&=\frac{\Gamma(1+\alpha)}{(1 - t\, x)^{1+\alpha}} \left[1 - \frac{t\, y \,c_{z}}{1 - t\,x}\right]^{-1-\alpha} \frac{1}{\Gamma(1+z)}\Big\vert_{z=\alpha} \nonumber \\
	&= \frac{1}{(1-t\,x)^{1+\alpha}} \left[\sum_{n=0}^{\infty} \left(\frac{t \,y}{1-t\,x}\right)^n \frac{\Gamma(1+\alpha+n)}{n!}\,  c^{n}_{z}\right] \frac{1}{\Gamma(1+z)}\Big\vert_{z=\alpha} \nonumber \\
	&= \frac{1}{(1-t\,x)^{1+\alpha}} \sum_{n=0}^{\infty} \frac{1}{n!}\left(\frac{t\, y}{1-t\,x}\right)^n \nonumber \\ 
	&= \frac{1}{(1-t\,x)^{1+\alpha}}\exp\left(\frac{t\,y}{1-t\,x}\right),
\end{align}
which for {$L_{n}^{(\alpha)}(y) = L_{n}^{(\alpha)}(1, -y)$} gives the well-known formula (8.975.1) on p.~1002 of \cite{ISGradshteyn07}. Furthermore, repeating a similar 
calculation for exponential generating function for integer $\alpha = m$ and $\beta=1$, we have
\begin{align}\label{EQ1.12}
	{_{E}G}_{1}^{(m, 1)}(x, y\,|\, t) &= \left[\sum_{n=0}^{\infty} \frac{(n + m)!}{(n!)^2} [t \,(x + y\, c_{z})]^n\right] \frac{1}{\Gamma(1+z)}\Big\vert_{z=m} \nonumber\\
	&=\left[ \E^{t\, (x + y \,c_{z})} \sum_{r=0}^{m} \binom{m}{r} \,\frac{m!}{r!} \,[t\,(x + y\, c_{z})]^r \right] \frac{1}{\Gamma(1+z)}\Big\vert_{z=m} \nonumber\\
	&= \E^{t x} \sum_{r=0}^{m} \binom{m}{r}\, \frac{m!}{r!} \left[\sum_{s=0}^{r} \binom{r}{s} \,(t \,x)^{r-s} \,(t \,y)^s \,{c}_{z}^{s} \E^{t\, y\, c_{z}}\right] \frac{1}{\Gamma(1+z)}\Big\vert_{z=m} \nonumber \\
	&= \E^{t x} \sum_{r=0}^{m} \binom{m}{r} \,\frac{m!}{r!} \,\left[\sum_{s=0}^{r} \binom{r}{s} \,(t\, x)^{r-s} \,(t \, y)^s \E^{t\, y\, c_{z}}\right] \frac{1}{\Gamma(1+z)}\Big\vert_{z=m+s} \nonumber \\
	&= \E^{t x} \sum_{r=0}^{m} \binom{m}{r} \,\frac{m!}{r!} \sum_{s=0}^{r} \binom{r}{s}\, (t\, x)^{r-s} \,(t\, y)^s \,W^{(1, m+s+1)}(t\, y), 
\end{align}
where we used the first Kummer relation for confluent hypergeometric series, see Eq.~(9.212) on p.~1023 of \cite{ISGradshteyn07}, in passing from the first to the second line.

\section{Lacunary generating functions of Laguerre polynomials}

In this paragraph we apply our method to treat various lacunary generating functions for the Laguerre polynomials. The formulas thus obtained appear, for the most part, to be new.

The double-lacunary exponential generating function of ordinary Laguerre polynomials $(\alpha=0)$
\begin{equation}\label{EQ2.1}
	{_{E}G}_{2}^{(1, 1)}(x, y\,|\, t) = \sum_{n=0}^{\infty} \frac{t^{n}}{n!}\, L_{2 n}(x, y)
\end{equation}
is apparently not known. In the following we obtain the explicit form of the series in 
Eq.~\eqref{EQ2.1} in terms of known functions.
According to our procedure, we rewrite 
Eq.~\eqref{EQ2.1} using Eq.~\eqref{EQ1.4} as 
\begin{align}\label{EQ2.2}
	{_{E}G}_{2}^{(1, 1)}(x, y\,|\, t) &= \left[\sum_{n=0}^{\infty} \frac{t^n}{n!}\, (x + y \, c_{z})^{2 n}\right] \frac{1}{\Gamma(1+z)}\Big\vert_{z=0}\nonumber\\ 
	&= \E^{t\,x^2} \exp\left[t\, y^2\,  c_z^2 + 2 \,t\,x\, y\,  c_{z}\right] \frac{1}{\Gamma(1+z)}\Big\vert_{z=0}.
\end{align}
We can provide a definite meaning for the previous expression in terms of known special functions by recalling the expansion
\begin{equation}\label{EQ2.3}
	\E^{b \,z^2 + a \, z} = \sum_{n=0}^{\infty} \frac{z^n}{n!}\, H_{n}^{(2)}(a, b), 
\end{equation}
with
\begin{equation}\nonumber
	H_{n}^{(2)}(a, b) = n! \sum_{r=0}^{\lfloor n/2\rfloor} \frac{a^{n - 2r}\, b^r}{(n - 2 r)! \, r!},
\end{equation}
where $\lfloor n\rfloor$ is the floor function. The polynomials $H_{n}^{(2)}(a, b)$ are two-variable Hermite polynomials \cite{PAppell29, DCocolicchio99, GDattoli04}, also defined through the operational rule given in Eq.~\eqref{30/10/2014-3}. They reduce to the ordinary Hermite polynomials $H_{n}(z)$ through the relation \eqref{30/10/2014-4}. The above family of polynomials provides a basis for the definition of the so-called Hermite-based ($H$-based) functions \cite{GDattoli00}; for example the Hermite-based ($H$-based) cylindrical Bessel functions are defined as
\begin{equation*}
	_{H}J_{n}(x, y) = \sum_{r=0}^{\infty}(-1)^r  \frac{H_{n + 2 r}^{(2)}(x, y)}{2^{n + 2r} \, r!\, (n+r)!}.
\end{equation*}
They have been obtained by replacing $x^{n + 2 r}$ in the relevant series expansion by $H_{n+2r}^{(2)}(x, y)$.

According to Eq.~\eqref{EQ2.3}, we obtain
\begin{align}
	\exp\left[{t\, y^2\, c_z^2 + 2 \,t\,x\, y\, c_z}\right] \frac{1}{\Gamma(1+z)}\Big\vert_{z=0} &= \left[\sum_{r=0}^{\infty} \frac{c_z^r}{r!}  H_{r}^{(2)}(2\,t\,x\,y , t\,y^2)\right]\frac{1}{\Gamma(1+z)}\Big\vert_{z=0} \nonumber\\ 
	&= \sum_{r=0}^{\infty} \frac{1}{(r!)^2} H_{r}^{(2)}(2\,t\, x\, y ,t\, y^2 ) \nonumber\\
	&= \sum_{r=0}^{\infty} \frac{(i\,\sqrt{t}\, y )^r}{(r!)^2} H_{r}(i \,\sqrt{t} \,x ) \label{EQ2.5} \\
	&= {_{H}C}_{0}(-2\, t\, x \,y, \, t\,y^{2}), \nonumber
\end{align}
where 
\begin{equation}\nonumber
	{_{H}C}_{\alpha}(x, y) = \sum_{r=0}^{\infty} \frac{H_{r}^{(2)}(-x, y)}{r!\,  \Gamma(1 + \alpha + r)}
\end{equation}
is the Hermite-based ($H$-based) version of the Bessel--Wright function
\begin{equation}\nonumber
	C_{\lambda}(x) = \sum_{r=0}^{\infty} \frac{(-x)^r}{r! \, \Gamma(1 + \lambda + r)} = c_{z}^{\lambda} \exp\left[{-x c_{z}} \right]\frac{1}{\Gamma(1+z)}\Big\vert_{z=0},
\end{equation}
{which goes over to the Bessel--Wright function $W^{\beta, \alpha}(x)$ for $\beta = 1$ and $\alpha = 1+\lambda$.} Thus we get in conclusion
\begin{equation*}
	{_{E}G}_{2}^{(1, 1)}(x, y\,|\, t) = \E^{x^2 \,t} {_{H}C}_{0}(-2\, x\, y\, t, y^2\, t).
\end{equation*}

In terms of standard Laguerre polynomials {$L_{n}(y) = L_{n}(1, -y)$} Eq.~\eqref{EQ2.1} can thus be rewritten as 
\begin{equation}\label{EQ2.7}
	\sum_{n=0}^{\infty} \frac{t^n}{n!}  L_{2n}(y) = \E^t {\sum_{r=0}^{\infty} \frac{(i\,\sqrt{t}\,y)^r}{(r!)^2}  H_{r}(i \sqrt{t})},
\end{equation}
which constitutes one of key results of the present investigation. The validity of identity \eqref{EQ2.7} as well as of further results of this type for other generating functions (see Eqs.~\eqref{EQ1.12}, and \eqref{EQ2.8}, \eqref{EQ2.14}, \eqref{EQ3.1}, \eqref{EQ3.3} below) can be independently proven by the substitution $t\to -t^{2}$, followed by coefficient extraction and the use of Eqs.~\eqref{EQ1.11} and \eqref{EQ2.3} \cite{Bostan}. 

In fact, it turns out that Eq.~\eqref{EQ2.7} is the special case of a more general relation which extends this result to the 
generalized Laguerre polynomials {$L^{(\alpha)}_{2n}(y)$}:
\begin{equation}\label{24/01/2015-1}
	\sum_{n=0}^{\infty} \frac{(\frac{1}{2})_{n}\, (-t^2)^n}{(\frac{\alpha}{2} + \frac{1}{2})_{n}\, (\frac{\alpha}{2}+1)_{n}} {L^{(\alpha)}_{2n}(y)} = \E^{-t^2}\sum_{n=0}^{\infty} \frac{{(y\,t)^n}}{n! \, (1+\alpha)_{n}} H_{n}(t).
\end{equation}
Using the shift operator $c_{z}$ of Eq.~\eqref{11/12/2014-1}, the demonstration of Eq.~\eqref{24/01/2015-1} works as follows:
\begin{align*}
	\sum_{n=0}^{\infty} \frac{(-t^2)^n}{n!}\frac{(2n)!}{\Gamma(1+\alpha+2n)} {L^{(\alpha)}_{2n}(y)} &= \left[\sum_{n=0}^{\infty} \frac{(-t^2)^n\, {(1-y \,c_z)^{2n}}}{n!}\right] \frac{1}{\Gamma(1+z)}\Big\vert_{z={\alpha}} \\
	& = \exp\left[{-t^2\, (1 - y\, c_{z})^2}\right] \frac{1}{\Gamma(1+z)}\Big\vert_{z=\alpha}\\ 
	&= \E^{-t^2} \sum_{n=0}^{\infty} \frac{H_{n}(t)}{n!} \left[t\, y\,c_{z}\right]^n \frac{1}{\Gamma(1+z)}\Big\vert_{z=\alpha} \\
	& = \E^{-t^2} \sum_{n=0}^{\infty} \frac{(t\,y)^n}{n! \,\Gamma(1+\alpha+n)} H_{n}(t).
\end{align*}
Note that Eq.~\eqref{24/01/2015-1} for $\alpha = 0$ and after substitution $-t^2 \rightarrow t$, reproduces Eq.~\eqref{EQ2.7}. We note for completeness another version of the identity Eq.~\eqref{24/01/2015-1} rewritten in terms of hypergeometric representations of Laguerre and Hermite polynomials \cite{APPrudnikov98}:
\begin{align*}
	& \sum_{n=0}^{\infty} \frac{\left(1/2\right)_{n} (1+\alpha)_{2n} \,(-t^2)^n}{(2n)! \,\left(\alpha/2 +1/2\right)_n \left(\alpha/2+1\right)_n} {_{1}F_{1}}\left({\genfrac {}{}{0pt}{}{-2\,n} {1 +\alpha}}\,;\,x\right) \\
	& \qquad\qquad\qquad = \E^{-t^2} \sum_{n=0}^{\infty} \frac{(2\, t^2 \, x)^n}{n! \, (1+\alpha)_n} {_{2}F_{0}}\left({\genfrac {}{}{0pt}{}{-n/2, (1-n)/2} - }\,;\, -\frac{1}{t^2}\right).
\end{align*}

Using the above technique, see Eqs.~\eqref{EQ2.1}-\eqref{EQ2.5}, we derive the exponential double-lacunary generating function for the generalized Laguerre polynomials $L_{n}^{(\alpha)}(x)$, for $\alpha~=~m~= ~1, 2, \ldots$ and establish that
\begin{equation*}
	\sum_{n=0}^{\infty} \frac{t^n}{n!} L_{2 n}^{(m)}(x, y) = \E^{t \,{x}^2} \sum_{r=0}^{\infty} \frac{p_{2 m}(r\,; x, y, t)}{r! \,(r + 3 m)!} {H^{(2)}_{r}(2\,t\, x\, y , t\, y^2)},
\end{equation*}
which for standard generalized Laguerre polynomials {$L^{(m)}_{n}(y) = L^{(m)}_{n}(1, -y)$}, using Eq.~\eqref{30/10/2014-4}, has the form
\begin{equation}\label{EQ2.8}
	\sum_{n=0}^{\infty} \frac{t^n}{n!} L_{2 n}^{(m)}({y}) = \E^{t} \sum_{r=0}^{\infty} \frac{p_{2 m}(r\,; {1, -y}, t)}{r!\, (r + 3 m)!} ({-i } \sqrt{t}\, y)^r H_{r}(i \sqrt{t}).
\end{equation}
The polynomials $p_{2 m}(r\,; x, y, t)$ are of degree $2 m$ in the variable $r$, with the coefficients depending on $x$, $y$ and $t$. For $m = 1, 2$, $p_{2 m}(r\,; x, y, t)$ have the explicit form
\begin{align}
	p_{2}(r\,; x, y, t) & = [1 + 2\, {x}^2 t] \,r^2 + [5  +  4\, x \,y\, t + 10 \,{x}^2 t]\, r \nonumber \\
	& \quad + [6 + 12 {x}^2 t+ 12 x\, y \,t + 2 {y}^2t ], \label{15/01/2015-4} \\
	p_{4}(r\,; x, y, t) & = [2 + 10  \,{x}^{2} t+ 4 \,{x}^{4}t^{2} ] \,r^{4}\nonumber\\
	&\quad + [36 + (180\, {x}^{2} + 20\, x\,y)\, t + (72 \,{x}^{4} + 16\, {x}^{3} {y})\,t^{2}]\, r^{3} \nonumber \\
	&\quad + [238 + (10\, {y}^{2} + 1190 \,{x}^{2} + 300 \, x\,y) \,t \nonumber\\
	&\quad\quad + (240 \,{x}^{3} {y} + 24 \,{x}^{2} {y}^{2} + 476\, {x}^{4})\, t^{2}]\, r^{2} \nonumber \\
	&\quad + [684 + (110 \,{y}^{2} + 1480 \, x\,y + 3420\, {x}^{2})\, t \nonumber \\
	&\quad \quad + (1184\, \,{x}^{3} {y} + 264\,  x^{2} y^{2}+ 16 \,x\, y^{3} + 1368\, {x}^{4}) \,t^{2}] \,r \nonumber \\
	&\quad + [720 + (2400 \,x\,y + 3600 \,{x}^{2} + 300 \,{y}^{2}) \,t  \nonumber \\
	&\quad \quad + (1920\, {x}^{3} {y} + 96 \,x \,y^{3} + 720\, x^{2} y^{2} + 1440 \,{x}^{4} + 4 \,{y}^{4})\, t^{2}]. \label{15/01/2015-5}
\end{align}
With a moderate effort the polynomials $p_{2m}(r; x, y, t)$ for $m > 2$ can be also obtained. Explicit forms, various generating functions and combinatorial interpretations of the polynomials \eqref{15/01/2015-4} and \eqref{15/01/2015-5} were recently obtained by Strehl \cite{Strehl, STR}. 

The current method can be extended to the more general cases like
\begin{equation}\nonumber
	{_{E}G}_{2}^{(\alpha, \beta)}(x, y\,|\, t) = \sum_{n=0}^{\infty} \frac{t^n}{n!}  \Lambda_{2 n}^{(\alpha, \beta)}(x, y),
\end{equation}
which, according to our procedure, can be written as
\begin{align}\label{EQ2.9}
	{_{E}G}_{2}^{(\alpha, \beta)}(x, y\,|\, t) &= \sum_{r=0}^{\infty} \frac{t^n}{n!}  (x + y\,c_z^{\beta})^{2n} \frac{1}{\Gamma(1+z)}\Big\vert_{z=\alpha} \nonumber \\
	& = \E^{x^2\, t} \exp\left[2\,t\,  x \,y\, c_z^{\beta} +  t \, y^2 \,c_z^{2 \beta}\right] \frac{1}{\Gamma(1+z)}\Big\vert_{z=\alpha}  \nonumber \\
	&= \E^{x^2\, t}\left[ \sum_{r=0}^{\infty} \frac{c_z^{\beta r}}{r!}  H^{(2)}_{r}(2\, t\, x\, y, t\, y^2)\right] \frac{1}{\Gamma(1+z)}\Big\vert_{z=\alpha} \nonumber \\
	&= \E^{x^2 t} {_{H}W}^{(\beta, \alpha + 1)}(2\, t\, x\, y, t\,y^2)
\end{align}
with
\begin{equation}\nonumber
	_{H}W^{(\alpha, \beta)}(x, y) = \sum_{r=0}^{\infty} \frac{H^{(2)}_{r}(x, y)}{r!  \,\Gamma(\alpha\, r + \beta)}
\end{equation}
being the Hermite-based ($H$-based) version of the Bessel--Wright function \cite{GDattoli00}.

We give here without demonstration the following two formulas of similar type
\begin{equation}\label{EQ2.10}
	\sum_{n=0}^{\infty} t^{n} \,L_{2n}(x) = \frac{1}{1-t} \sum_{r=0}^{\infty} \frac{L_{r}^{(r)}(x/2)}{(1/2)_{r}} \left[-\frac{t\,x}{2\,(1-t)}\right]^{r}
\end{equation}
and
\begin{equation}\label{EQ2.11}
	\sum_{n=0}^{\infty} t^{n}\, L_{3n}(x) = \frac{1}{1-t} \sum_{r=0}^{\infty} \left(-\frac{3\,t\,x}{1-t}\right)^{r} \left[\sum_{s=0}^{r} \frac{r!\, (-x)^{s}}{(r-s)!\, (r + 2s)!} L_{s}^{(s+r)} \left({x/3}\right)\right].
\end{equation}

In this section we have shown that the tools of employing the symbolic method to derive lacunary generating functions for Laguerre polynomials can easily be applied with a minimum of computation effort.

We now employ the symbolic method to derive old and further new ordinary generating functions for the generalized Laguerre polynomials. We start with a simple observation, namely
\begin{equation}
\label{EQ2.12}
	c_{z}^{\alpha} \exp\left[{b\, c_{z}^{-1}}\right] \frac{1}{\Gamma(1+z)}\Big\vert_{z=0} = \left[\sum_{n=0}^{\infty} \frac{b^n}{n!} c_{z}^{\alpha-n}\right]\frac{1}{\Gamma(1+z)}\Big\vert_{z=0} = \frac{(1 + b)^{\alpha}}{\Gamma(1+\alpha)}.
\end{equation}
See also Ref.~\cite{DBabusci13} for a  demonstration. 
We calculate the generating function 
$\mathcal{G}^{(\alpha)}_{1}(x, y\,\vert\, t) =  \sum_{n=0}^{\infty} t^n \, L_{n}^{(\alpha - n)}(x, y)$, which after applying Eqs.~(\ref{12/12/2014-1}), (\ref{11/12/2014-3aa}) and (\ref{EQ2.12}) can be written as
\begin{align}\label{EQ2.13}
	\mathcal{G}^{(\alpha)}_{1}(x, y\,\vert\, t) &=  \Gamma(1+\alpha) \left[c_{z}^{\alpha}\sum_{n=0}^{\infty} \frac{t^n}{n!} \left(\frac{{x + y \,c_{z}}}{c_{z}}\right)^n\right] \frac{1}{\Gamma(1+z)}\Big\vert_{z=0}  \nonumber\\
	&= \Gamma(1+\alpha)\, \E^{{t \,y}} \,c_{z}^{\alpha}\, \exp\left[{t} \,x\, c_{z}^{-1}\right]\, \frac{1}{\Gamma(1+z)}\Big\vert_{z=0} = (1+ {t} \,x)^{\alpha} \E^{t \,{y}}, 
\end{align}
Formula \eqref{EQ2.13} for the ordinary Laguerre polynomials, i. e. for $y=1$, is equal to (5.11.4.8) on p.~706 of \cite{APPrudnikov98}. In the case of generating function
\begin{equation}\label{27/07-1}
	\mathcal{G}^{(\alpha)}_{2}(x, y\,\vert\, t) = \sum_{n=0}^{\infty} t^{n}\,  L_{2 n}^{(\alpha - 2 n)}(x, y) 
\end{equation}
Eqs.~(\ref{12/12/2014-1}) and (\ref{11/12/2014-3c}) yield
\begin{align}
\label{EQ2.14}
	\mathcal{G}^{(\alpha)}_{2}(x, y\,\vert\, t) &=   \Gamma(1+\alpha) \left[c_{z}^{\alpha}\sum_{n=0}^{\infty} \frac{t^{n}}{(2 n)!}  \left(\frac{x + y\,c_{z}}{c_{z}}\right)^{2 n}\right] \frac{1}{\Gamma(1+z)}\Big\vert_{z=0} \nonumber \\
	&=   \Gamma(1+\alpha)\left[c_{z}^{\alpha} \cosh\left(\sqrt{t}\,\frac{x}{c_{z}} + \sqrt{t}\,y \right) \right]\frac{1}{\Gamma(1+z)}\Big\vert_{z=0} \nonumber\\
	&=   \left[\cosh(\sqrt{t}\,{y} ) \cosh\left(\sqrt{t}\,\frac{x}{c_{z}}\right) {+} \sinh(\sqrt{t}\,{y}) \sinh\left(\sqrt{t}\,\frac{x}{c_{z}}\right)\right]  \frac{\Gamma(1+\alpha)}{\Gamma(1+z)}\Big\vert_{z=\alpha} \nonumber \\
	& = (1 - t\, x^{2})^{
	{\alpha/2}} 
	\left\{\cosh(\sqrt{t}\,y)
	\cosh[i\,T_{\alpha}(t, x)] 
	+ \sinh(\sqrt{t}\,y)
	\sinh[i\,T_{\alpha}(t, x)]\right\} 
	\nonumber \\
	& = (1 - t\, x^{2})^{
	{\alpha/2}} 
	\cosh[\!\sqrt{t}\, y 
	+ i \,T_{\alpha}(t, x)],
%
\end{align}
where
\begin{equation}\nonumber
	T_{\alpha}(t, x) = \alpha\arcsin\left(\frac{\sqrt{t}\,{x}}{\sqrt{t\,{x}^2 -1}}\right).
\end{equation}

\noindent
\textit{Remark:} 
The use of the initial definition of $\Lambda_{n}^{(\alpha)}(x, y)$, see Eqs.~\eqref{11/12/2014-3c} and \eqref{11/12/2014-3aa}, in the treatment of $\mathcal{G}_{1}^{(\alpha)}(x, y\,|\, t)$ and $\mathcal{G}_{2}^{(\alpha)}(x, y\,|\, t)$ leads to considerable difficulties in obtaining the final results of Eqs.~\eqref{EQ2.13} and \eqref{EQ2.14}. One way to avoid these difficulties is to adopt a modified (albeit strictly equivalent) definition of $\Lambda_{n}^{(\alpha-n)}(x, y)$ in the form 
\begin{equation*}
	\Lambda_{n}^{(\alpha-n)}(x, y) = c_{z}^{\alpha-n}(x+y\,c_{z})^{n} \frac{1}{\Gamma(1+z)}\big\vert_{z=0}.
\end{equation*}
Actually this last definition was used in deriving Eqs.~\eqref{EQ2.13} and \eqref{EQ2.14}, whereas the previous one was used everywhere else.

\section{Further developments}

The method we have developed so far can be extended to obtain even slightly more complicated expressions like
\begin{equation}\nonumber
	{_{E}G}_{2, \ell}^{(1, 1)}(x, y\,|\, t) = \sum_{n=0}^{\infty} \frac{t^n}{n!} L_{2 n + \ell}(x, y) \quad \text{and} \quad {_{E}G}_{3, \ell}^{(1, 1)}(x, y\,|\, t) = \sum_{n=0}^{\infty} \frac{t^n}{n!}  L_{3n + \ell}(x, y),
\end{equation}
for $l = 1, 2, \ldots$. Note that corresponding formulas for the Hermite polynomials were obtained in \cite{IMGessel05}. Applying Eqs.~\eqref{11/12/2014-2a}, \eqref{EQ2.2}, and \eqref{EQ2.3} for the double lacunary generating function of Laguerre polynomials,
we have
\begin{align*}
	{_{E}G}_{2, \ell}^{(1, 1)}(x, y\,|\, t) & = \E^{t\,{x}^2} \left[({x + c_{z} y})^\ell \sum_{r=0}^{\infty} \frac{c_{z}^r}{r!}\,  H^{(2)}_{r}(2\, t\, x\, y, t\,{y}^2)\right] \frac{1}{\Gamma(1+z)}\Big\vert_{z=0} \\
	& = \E^{t\,{x}^2} \sum_{s=0}^{\ell} \binom{\ell}{s} \,{x}^{\ell-s} {y}^s \,_{H}C_{s}({-}2 \,t\,x \,y , t\,{y}^2),
\end{align*}
which, for $L_{n}(y) = L_{n}(1, -y)$,can be written in terms of standard generalized Laguerre and Hermite polynomials as follows:
\begin{align}
\label{EQ3.1}
	\sum_{n=0}^{\infty} \frac{t^n}{n!}L_{2 n + \ell}(x) &= \E^{t} \sum_{r=0}^{\infty} \frac{({-i \sqrt{t}})^r}{r!} H_{r}(i\sqrt{t}\, {x}) \sum_{s=0}^{\ell} \binom{\ell}{s} \frac{x^{{l-s}}}{(r+s)!} \nonumber \\
	&= \E^t \ell! \sum_{r=0}^{\infty} \frac{ (i\sqrt{t}\, x )^r}{r! \,(\ell+r)!} \,L_{\ell}^{(r)}(x)\, H_{r}(i \sqrt{t}).
\end{align}
Using the analogous procedure, we obtain also ${_{E}G}^{(1, 1)}_{3, \ell}(x, y \,|\,t)$ in the form
\begin{align}\label{EQ3.2}
	{_{E}G}_{3, \ell}^{(1, 1)}(x, y\,|\, t) &= \E^{t\,x^3}\, \left[(x + y \,c_{z} )^{\ell} \sum_{n=0}^{\infty} \frac{c_{z}^n}{n!}\,  H_{n}^{(3)}(3\, t  \,x^2\,y, 3\,t\, x\, y^2, t\,y^3 )\right] \frac{1}{\Gamma(1+z)}\Big\vert_{z=0} \nonumber \\
	&= \E^{t\,x^3} \sum_{s=0}^{\ell} \binom{\ell}{s}\,  y^s x^{\ell-s}{}  _{H}C_{s}^{(3)}(3\, t\,x^2 \,y , 3\, t \,x \,y^2 , t\,y^3),
\end{align}
where
\begin{equation*}
	_{H}C_{s}^{(3)}(x, y, z) = \sum_{r=0}^{\infty} \frac{H_{r}^{(3)}(x, y, z)}{r!  \,(r+s)!}
\end{equation*}
is a third order Hermite-based ($H$-based) Tricomi function \cite{GDattoli00}, with
\begin{equation*}
	H_{n}^{(3)}(x, y, z) = n! \sum_{r=0}^{\lfloor n/3\rfloor} \frac{z^r \,H^{(2)}_{{n - 3 r}}(x, y)}{r!\,(n - 3 r)!}
\end{equation*}
being a third order three-variable Hermite polynomial, with the generating function \cite{GDattoli02}
\begin{equation*}
	\sum_{n=0}^{\infty} \frac{t^n}{n!}  H_{n}^{(3)}(x, y, z) = \E^{t\,x  + t^2\,y  + t^3\,z }.
\end{equation*}
Eq.~\eqref{EQ3.2} expressed via standard Laguerre and Hermite polynomials is given as
\begin{equation}
\label{EQ3.3}
	\sum_{n=0}^{\infty} \frac{t^n}{n!} L_{3 n + \ell}(x) = \E^t \ell! \sum_{n=0}^{\infty} \frac{L_{\ell}^{(n)}(x)}{(n+\ell)!} \sum_{r=0}^{\lfloor n/3\rfloor} \frac{(-t\,x^3)^r (i \sqrt{3\, t}\,x )^{n - 3 r}}{r! \, (n-3\,r)!} H_{n - 3r}\left(i\frac{\sqrt{3\,t}}{2}\right).
\end{equation}
In the case of the generalized 
Laguerre polynomials for $\alpha=1$ and $\ell=0$, $L_{3 n}^{(1)}({1, -y}) = L_{3 n}^{(1)}(y)$, the generating function is given as
\begin{align}
\label{EQ3.4}
	&\sum_{n=0}^{\infty} 
	\frac{t^{n}}{n!} 
	L^{(1)}_{3 n}(y) = 
	\E^{t} \sum_{r=0}^{\infty} 
	\frac{q_{3}(r\,; y, t)}{r!\, (r + 4)!} 
	H^{(3)}_{r}
	(- 3 \,t \,y, 3 \,t\, y^{2}, - t\, y^{3}), \nonumber\\
	&\qquad = \E^{t} 
	\sum_{r=0}^{\infty} 
	\frac{y^r\,q_{3}(r\,; y, t)}{ (r + 4)!} 
	\sum_{s=0}^{\lfloor r/3 \rfloor} 
	\frac{(-t )^s 
	(i\sqrt{3\,t} )^{r-3\, s}}{s! \, (r-3s)!} H_{r - 3s}
	\left(i \frac{\sqrt{3 \,t}}{2}\right)
\end{align}
with
\begin{align*}
	q_{3}(r\,; y, t) &= 
	[1 + 3 \,t]\, r^{3} 
	+ [9 + 27\, t - 9 \, y\,t]\, r^{2} 
	+ [26 + 78 \,t - 63\, y\,t + 9\, y^{2}\,t]\, r \\
	&\quad+ [24 + 72 \,t - 108 \,y\,t + 36  y^{2} \,t- 3\, y^{3}\,t],
\end{align*}
which is a third order polynomial in $r$, with the coefficients depending on $y$ and $t$. Its closed form and the analytical and combinatorial properties were worked out by Strehl \cite{Strehl, STR}.

The extension to the case
\begin{equation}\nonumber
	{_{E}G}_{m, \ell}^{(1, 1)}(x, y\,|\, t) = \sum_{n=0}^{\infty} \frac{t^n}{n!}  L_{m n + \ell}(x, y)
\end{equation}
can be straightforwardly accomplished within the present framework and reads
\begin{align}\label{EQ3.5}
	{_{E}G}_{m, \ell}^{(1, 1)}(x, y\,|\, t) &= \E^{t \,y^m} \left[(x + y\, c_{z} )^\ell \sum_{r=0}^{\infty} \frac{c_{z}^r}{r!} \,H_{r}^{(m)}(-\alpha_{1}, \alpha_{2}, \ldots, (-1)^m \alpha_{m})\right] \frac{1}{\Gamma(1+z)}\Big\vert_{z=0} \nonumber \\
	&= \E^{t\, x^m} \sum_{s=0}^{\ell} \binom{\ell}{s}\, x^{\ell-s} \,y^{s} \,{_{H}C}^{(m)}_{s}(\alpha_{1}, \alpha_{2}, \ldots, \alpha_{m}), \nonumber \\ \alpha_{p} 
	&= \binom{m}{p}\, x^{m-p}\, (-y)^p \,t, \quad p=1, 2, \ldots m, 
\end{align}
where $_{H}C_{n}^{(m)}\left(\alpha_{1}, \alpha_{2}, \ldots, \alpha_{m}\right)$ is an $m$-th order Hermite-based Tricomi function, defining as basis Hermite polynomials $H_{n}^{(m)}(-\alpha_{1}, \alpha_{2}, \ldots, (-1)^m \alpha_{m})$ in $m$ variables, as specified by the generating function \cite{GDattoli02}
\begin{equation*}
	\sum_{n=0}^{\infty} \frac{t^n}{n!}  H_{n}^{(m)} \left(x_{1}, x_{2}, \ldots, x_{m}\right) = \exp\left(\sum_{s=1}^{m} x_{s} \,t^s\right).
\end{equation*}

Let us finally consider the bilateral generating function
\begin{multline*}
	\sum_{n=0}^{\infty} \frac{t^n}{n!}  L_{n}(x, y) \, L_{n}(v, u) = \\ \sum_{n=0}^{\infty} \frac{t^n}{n!} \left[(x + y\,c_{z} )^n \frac{1}{\Gamma(1+z)}\Big\vert_{z=0}\right]\,\left[(v +  u\,c_{\tilde{z}})^{n} \frac{1}{\Gamma(1+\tilde{z})}\Big\vert_{\tilde{z}=0}\right].
\end{multline*}
We obtain
\begin{align*}
	\sum_{n=0}^{\infty} \frac{t^n}{n!}  L_{n}(x, y)  \,L_{n}(v, u) &= \E^{t\,x\, v} \left[\E^{- t\,y \, v \, c_{z}} \frac{1}{\Gamma(1+z)}\Big\vert_{z=0}\right]\, \left[\E^{- t \,x \, u\, c_{\tilde{z}}}\frac{1}{\Gamma(1+\tilde{z})}\Big\vert_{\tilde{z}=0}\right]\nonumber\\
	&\qquad\times \left(\E^{t\,y\,u\,  c_{z}\, c_{\tilde{z}}} \frac{1}{\Gamma(1+z)}\Big\vert_{z=0}\, \frac{1}{\Gamma(1+{\tilde{z}})}\Big\vert_{{\tilde{z}}=0}\right) \nonumber\\
	&= \E^{t\,x\,v\, t}  {_{H}C_{0, 0}}(t\,y\,v, t\,x\,u \,|\,t\, y\,u),
\end{align*}
where 
\begin{equation*}
	_{H}C_{0, 0}(x, y\,|\, \tau) = \sum_{r, s, k = 0}^{\infty} \frac{x^{r}\, y^{s}\, \tau^{k}}{r! \,s! \,k! \,(r + k)!\,(s + k)!}.
\end{equation*}
Closely related considerations devoted to combinatorics of Laguerre polynomials are developed in \cite{DFoata84}.

Using the operational method, we can also obtain other interesting identities for the lacunary
generating functions of the  generalized Laguerre polynomials. These formulas are listed below:
\begin{align}\label{EQ3.9}
	\sum_{n=0}^{\infty} \frac{\left(\frac{1}{2}\right)_{n} t^n}{\left(1+\frac{\alpha}{2}\right)_n} L_{2n}^{(\alpha)}(x) 
	&= (1-t)^{-(1+\alpha/2)}\sum_{r=0}^{\infty} \frac{L_{r}^{(r+\alpha)}\ \left({x/2}\right)}{\left(1+\alpha/2\right)_r} \left(-\frac{t\,x}{2\,(1-t)}\right)^r \nonumber \\
	& = \frac{1}{\sqrt{1-t}} \left(\sqrt{t}\,\frac{x }{2}\right)^{-{\alpha/2}} \exp\left(-\frac{ t\,x}{1-t}\right) \,I_{\alpha/2}\left(\sqrt{t}\,\frac{x}{1-t}\right), 
\end{align}
obtained with formula (5.11.4.12) of \cite{APPrudnikov98}, which for $\alpha = 2m$ gives
\begin{align}\label{EQ3.10}
\sum_{n=0}^{\infty} \frac{\left(\frac{1}{2}\right)_{n}\,t^n}{(1+m)_{n}} L_{2 n}^{(2 m)}(x) 
	&= \frac{1}{\sqrt{1-t}} \left(\sqrt{t}\,\frac{x}{2}\right)^{-m} \exp\left(-\frac{t\,x}{1-t}\right) \,I_{m}\left(\sqrt{t}\,\frac{x}{1-t}\right), 
\end{align}
where $m=0, 1, 2, \ldots$ and $I_{m}(z)$ is the modified Bessel function;
\begin{multline}\label{EQ3.11}
	\sum_{n=0}^{\infty} \frac{\left(1/3\right)_{n} \,\left(2/3\right)_{n} t^n }{\left(1+\alpha/3\right)_{n}\, \left(2/3+\alpha/3\right)_{n}}\, L_{3 n}^{(\alpha)}(x) \cr 
	=(1-t)^{-(1+\alpha)/3} \sum_{r=0}^{\infty} \frac{\Gamma(1+\alpha+3r)}{\left(1+{\alpha/3}\right)_{r} \,\left({2/3}+{\alpha/3}\right)_{r}} \left(-\frac{t\,x}{9\,(1-t)}\right)^r \\
	\times \left[\sum_{s=0}^{r} \frac{(-x)^s \,L_{s}^{(\alpha+r+s)}\left({x/3}\right)}{(r-s)!\; \Gamma(1+\alpha+r+2s)}\right]. 
\end{multline}
Eq.~\eqref{EQ3.10} corrects the formula (5.11.2.10), p.~704 of \cite{APPrudnikov98}. 

\medskip

Let us now derive Eq.~\eqref{EQ3.9}. Following Eqs.~\eqref{12/12/2014-1}, \eqref{11/12/2014-3c} and the Gauss--Legendre formula for $\Gamma(2\,z)$, we get
\begin{align*}
	&\sum_{n=0}^{\infty} \frac{\left({1/2}\right)_{n} t^n}{\left({1+\alpha/2}\right)_n} L_{2n}^{(\alpha)}(x) = \sum_{n=0}^{\infty} \frac{\left({1/2}\right)_{n}\, t^n}{(2n)!}\,\frac{\Gamma(1+\alpha+2n)}{\left(1+\alpha/2\right)_{n}}\, \left[c_z^{\alpha}\, (1-x\,c_z)^{2n}\right]\,\frac{1}{\Gamma(1+z)}\Big\vert_{z=0} \\
	&\qquad = \Gamma(1+\alpha)\,\left[ c_z^{\alpha} \sum_{n=0}^{\infty} \frac{\left(1/2+\alpha/2\right)_{n}}{n!} [t\,(1-x\,c_z)^{2}]^n \right]\frac{1}{\Gamma(1+z)}\Big\vert_{z=0} \\
	&\qquad = \frac{\Gamma(1+\alpha) }{(1-t)^{(1+\alpha)/2}} \left[1 + c_z\left(\frac{2\,t\,x - t\,x^2 \,c_z}{1-t}\right)\right]^{-(1+\alpha)/2} \frac{1}{\Gamma(1+z)}\Big\vert_{z=\alpha} \\
	&\qquad = \frac{\Gamma(1+\alpha)}{(1-t)^{(1+\alpha)/2}}\sum_{r=0}^{\infty} \frac{(-1)^r}{r!} \left(\frac{1+\alpha}{2}\right)_{r}\, c_z^{r}\left(\frac{2\,t\,x- t\,x^2\, c_z}{1-t}\right)^r \frac{1}{\Gamma(1+z)}\Big\vert_{z=\alpha} \\
	&\qquad = \frac{\Gamma(1+\alpha)}{(1-t)^{(1+\alpha)/2}} \sum_{r=0}^{\infty} \frac{(-1)^r}{r!} \left(\frac{1+\alpha}{2}\right)_{r} \Lambda_{r}^{(r+\alpha)}\left(\frac{t\,x^2}{1-t}, \frac{2\,t\,x}{1-t}\right) \\
	&\qquad  = (1-t)^{-(1+\alpha)/2} \sum_{r=0}^{\infty} \frac{L_{r}^{(r+\alpha)}(x/2)}{\left(1+{\alpha/2}\right)_{r}} \left(-\frac{t\,x}{2\,(1-t)}\right)^r.
\end{align*}
Eq.~\eqref{EQ3.10} comes from Eq.~\eqref{EQ3.9} for $\alpha = 2m$ and using formula (5.11.4.12) on p.~706 of \cite{APPrudnikov98}. An alternative demonstration of Eq.~\eqref{EQ3.10} can be carried through using mixed bilateral generating function of Laguerre and Gegenbauer polynomials, see Appendix A.

\section{Concluding Remarks}

Having obtained the summation formulas Eqs.~\eqref{EQ3.9} and \eqref{EQ3.10}, we may write down the right hand sides of Eqs.~\eqref{2/03/2015-a} and \eqref{31/10/2014-7} in explicit form. For $\alpha = 0$, the initial condition for Eqs.~\eqref{31/10/2014-4} and \eqref{31/10/2014-5}, $f(x) = I_{0}(\beta x)$, $\beta > 0$, evolves with time according to  
\begin{align}\label{1/03/2015-1a}
	\exp[-t\,D_{x}^{(0)}]\, I_{0}(\beta x) & = \frac{1}{\sqrt{1 - \beta^2\, t^2}} \exp\left(-\frac{\beta^2\, t\,x}{1 - \beta^2 \,t^2}\right) \,I_{0}\left(\frac{\beta\, x}{1 - \beta^2 \,t^2}\right), \\
	& \equiv \mathcal{P}_{0}(t, x) \,I_{0}\left(\frac{\beta \,x}{1 - \beta^2\, t^2}\right), \quad (\beta \,t)^2 < 1. \label{1/03/2015-1b}
\end{align}
Furthermore, from Eq.~\eqref{1/03/2015-1a} one obtains
\begin{multline}\label{2/03/2015-1}
	\exp[-t\,D_{x}^{(\alpha)}] \,x^{-{\alpha/2}}\, I_{\alpha/2}(\beta \,x) \\ 
	= \frac{(1-\beta^2\,t^2)^{-(1+\alpha)/2}}{\Gamma(1+\alpha/2)} \exp\left(-\frac{\beta^2\, t \,x}{1-\beta^2\, t^2}\right) \left(\!\frac{x}{1-\beta^2\, t^2}\!\right)^{-\alpha/2} \,I_{\alpha/2}\left(\frac{\beta \,x}{1 - \beta^2 \,t^2}\right), \\ 
	\equiv \mathcal{P}_{\alpha}(t, x) \left(\frac{x}{1-\beta^2 \,t^2}\right)^{-{\alpha/2}} I_{{\alpha/2}}\left(\frac{\beta \,x}{1 - \beta^2 \,t^2}\right), \quad (\beta \,t)^2 < 1.
\end{multline}
In Eq.~\eqref{2/03/2015-1} we defined the ``prefunction" $\mathcal{P}_{\alpha}(t, x) > 0$ satisfying $\mathcal{P}_{\alpha}(0, x)~=~1$. Eqs.~\eqref{1/03/2015-1a} and \eqref{1/03/2015-1b} illustrate the scaling property of the time evolution: we show that special solutions exists which, up to the ``prefunction" $\mathcal{P}_{\alpha}(t, x)$, consist in rescaling of the argument of the initial condition $f(x)$ according to $x \to \beta x/(1 - \beta^2 \,t^2)$. This feature is characteristic for a number of examples of the so-called generalized Glaisher relations, see \cite{HOHE} and references therein. An elementary example of a relation of this type is given in Eq.~\eqref{31/10/2014-2}. Many similar examples can be worked out by choosing different forms of $f(x)$ and different values of $m$ in Eq.~\eqref{EQ3.10}.

Before concluding this paper it is  worth presenting some further comments to reconcile the present results with previous approaches based on the monomiality principle \cite{DCocolicchio99, DAT, GDattoli05a, GDattoli04, KAPenson09}. Within that context the so called Laguerre derivative ${D}^{(0)}_{x}$ has been introduced in Eq.~\eqref{30/10/2014-1}, so that 
\begin{equation*}
	D^{(0)}_{x} L_{n}(x, y) = n\, L_{n-1}(x, y).
\end{equation*}
The Laguerre derivative in differential terms is defined as $D^{(0)}_{x} = \frac{d}{dx} x\frac{d}{dx}$. 
As a further example of application we consider the action of the exponential containing the Laguerre derivative acting on the Tricomi function $C_{0}(x)~=~I_{0}(2\sqrt{x})$, namely
\begin{equation*}
	\exp[y\, D^{(0)}_{x}]\, C_{0}(x) = \ldots = \E^{y} C_{0}(x),
\end{equation*}
in agreement with the fact that $C_{0}(x)$ is an eigenfunction of the Laguerre derivative \cite{DCocolicchio99, GDattoli04} with the eigenvalue equals to one. An alternative operator definition of the Laguerre derivative is given in Appendix~B, where also some of its consequences are explored. The formal procedure we have developed can be pushed even further. 

The use of the following identity
\begin{equation*}
	K^{x \frac{d}{dx}} f(x) = f(K x).
\end{equation*}
and the definition of the 0-th order cylindrical Bessel function as a pseudo-Gaussian \cite{GDattoli00}, namely
\begin{equation*}
	J_{0}(x) = \E^{-c_z\, \left(x/{2}\right)^2} \frac{1}{\Gamma(1+z)}\Big\vert_{z=0},
\end{equation*}
allow for the derivation of the following identity
\begin{equation*}
	c_z^{- \frac{1}{2} x \frac{d}{d x}} J_{0} (x) = \E^{- \left(x/{2}\right)^2},
\end{equation*}
which, in terms of integral transforms, can be interpreted as a kind of Borel transform \cite{GDattoli05a}
\begin{equation*}
	c_z^{- \frac{1}{2} x \frac{d}{d 	x}} J_{0} (x) = \int_{0}^{\infty} \E^{-s} J_{0}\Big(\sqrt{s} x\Big)\, ds.
\end{equation*}
Finally let us note that, having expressed the cylindrical Bessel functions in terms of Gaussian, it is also possible to ``reduce" a Gaussian to a Lorentzian, according to the identity
\begin{equation*}
	\E^{-x^2} = \frac{1}{1 + x^2\,c_z} \frac{1}{\Gamma(1+z)}\Big\vert_{z=0}.
\end{equation*}
According to this last identity, we can write the relevant integral as
\begin{equation*}
	\int^{x}_{0} \E^{-\xi^2} d\xi = \int_{0}^{x} \frac{d\xi}{1+ c_z \xi^2} \frac{1}{\Gamma(1+z)}\Big\vert_{z=0} = \left[c_z^{-{1/2}} \arctan\Big(\sqrt{c_z}\, x \Big)\right] \frac{1}{\Gamma(1+z)}\Big\vert_{z=0}.
\end{equation*}
Here we give for completeness the list of new closed form expressions obtained in the present investigation. These are equations: \eqref{EQ1.7}, \eqref{EQ1.9}, \eqref{EQ1.12}, \eqref{EQ2.7}, \eqref{24/01/2015-1}, \eqref{EQ2.8}, \eqref{EQ2.9}, \eqref{EQ2.10}, \eqref{EQ2.11}, \eqref{EQ2.14}, \eqref{EQ3.1}, \eqref{EQ3.3}, \eqref{EQ3.4}, \eqref{EQ3.5}, \eqref{EQ3.9}, \eqref{EQ3.10}, and \eqref{EQ3.11}. The methods we have illustrated in this paper appear fairly flexible and amenable for further implementations, as will be shown in a future investigation.

\section{Acknowledgments}
\noindent
We thank Volker Strehl, Dominique Foata and Alin Bostan for important discussions. 

A short time after submission of the original version of the present article, Volker Strehl informed the authors that he has proven by purely combinatorial means virtually all the new summation formulas reported here. He has also provided explicit forms of polynomials $p_{2}(r\,; x, y, t)$, see Eq.~\eqref{15/01/2015-4}, $p_{4}(r\,; x, y, t)$, see Eq.~\eqref{15/01/2015-5}, and $q_{3}(r\,; x, t)$ on page~12, thereby closing the remaining gaps. In addition he has obtained a number of new formulas. 

We believe that Strehl's significant results constitute a very substantial step towards establishing new links between the umbral and combinatorial methods. We thank Dr. Strehl for encouragement and for ample correspondence on these subjects. We thank both referees for constructive remarks. 

This work has been supported by Agence Nationale de la Recherche (Paris, France) under Program PHYSCOMB No. ANR-08-BLAN-0243-2 and PHC Polonium, Campus France, project no. 28837QA. KG thanks for the support from MNiSW, Warsaw, Poland, ``Iuventus Plus 2015-2016", program no. IP2014 013073.

\appendix
\section{}
This appendix is entirely based on the material kindly communicated to the authors by Strehl \cite{Strehl1}.

The identity Eq.~\eqref{EQ3.10},
\begin{equation}\label{28/01/2015-1}
	\sum_{n=0}^{\infty} \frac{\left({1/2}\right)_{n}\, t^n}{(1+m)_n} L_{2n}^{(2m)}(x) = \frac{1}{\sqrt{1-t}}\left(\frac{\sqrt{t}\,x}{2}\right)^{-m} \exp\left(\frac{-t\,x}{1-t}\right) \, I_{m}\left(\frac{\sqrt{t}\,x}{1-t}\right),
\end{equation}
for $0 < t < 1$ and $m = 0, 1, \ldots$, will be shown to follow from Eq.~\eqref{EQ1.6}, \S~144, p.~281 of \cite{Rain}, namely from the following mixed bilateral generating function of products of Laguerre and Gegenbauer polynomials:
\begin{equation}
\label{28/01/2015-2}
	\sum_{n=0}^{\infty} \frac{n!\, t^n }{(\gamma)_{n}} L^{(\gamma-1)}_{n}(y) \,C^{{\gamma/2}}_{n}(x) = \rho^{-\gamma} \exp\left(\frac{-t\,(x-t)\,y}{\rho^2}\right) \,{_{0}F_{1}}\left({\genfrac {}{}{0pt}{}- { (1+\gamma)/2}}; \frac{t^2\,y^2}{4\, \rho^4}(x^2-1)\right),
\end{equation}
where $C^{{\gamma/2}}_{n}(x)$ is a Gegenbauer polynomials  and $\rho = \sqrt{1 -2xt + t^2}$. (As apparently first shown by Weisner \cite{Weis}, Eq.~\eqref{28/01/2015-2} can be derived using group-theoretic methods.)

On the left hand side of Eq.~\eqref{28/01/2015-2} we set $x=0$ and observe that $C^{\nu}_{2n}(0) = (-1)^n (\nu)_{n}/n!$ and $C^{\nu}_{2n+1}(0) = 0$, see p.~732 of \cite{APPrudnikov98}; then in the sum, only even terms in $n$ survive. Furthermore, upon replacing $t\rightarrow i\sqrt{t}$ in this sum we end up with
\begin{equation}\label{28/01/2015-3}
	\sum_{n=0}^{\infty} \frac{(2\,n)!}{(\gamma)_{2n}} \frac{(-1)^n}{n!} L^{(\gamma-1)}_{2n}(y)\, (i\sqrt{t})^{2n} \left({\gamma/2}\right)_n = \sum_{n=0}^{\infty} \frac{\left({1/2}\right)_n\,t^n }{\left({1/2+\gamma/2}\right)_n} L^{(\gamma-1)}_{2n}(y),
\end{equation} 
where we have used the relation
\begin{equation*}
	\frac{(2\,n)!}{(\gamma)_{2n}} \frac{\left({\gamma/2}\right)_n}{n!} = \frac{\left({1/2}\right)_n}{\left({(1+\gamma)/2}\right)_n}.
\end{equation*}
Observe that Eq.~\eqref{28/01/2015-3} can be identified with the left hand side of \eqref{28/01/2015-1} by setting $m=(\gamma-1)/2$. We apply now the above substitutions ($x=0$ and $t\rightarrow i\sqrt{t}$) to the right hand side of Eq.~\eqref{28/01/2015-2} which then becomes
\begin{equation}\label{28/01/2015-5}
	(1-t)^{-{\gamma/2}} \exp\left(-\frac{t\,y}{1-t}\right) \,{_{0}F_{1}}\left({\genfrac {}{}{0pt}{}-  {(1+\gamma)/2}}; \left(\frac{\sqrt{t}\,y }{2\,(1-t)}\right)^2\right).
\end{equation}
In the final step, the use in Eq.~\eqref{28/01/2015-5} of the hypergeometric representation of Bessel functions $I_{m}(z)$, see p.~729 of \cite{APPrudnikov98},
\begin{equation*}
	I_{m}(z) = \frac{1}{\Gamma(1+m)} \left(\frac{z}{2}\right)^m {_{0}F_{1}}\left({\genfrac {}{}{0pt}{}-  {1+m}}; \frac{z^2}{4}\right),
\end{equation*}
accomplishes the proof of Eqs.~\eqref{EQ3.10} and \eqref{28/01/2015-1}.

\section{}

In the main body of the paper we have introduced the concept of Laguerre derivative, which is a useful tool to deal with special polynomials of the Laguerre type. More generally it provides a key operator to further simplify many of the computational tasks we have dealt with, provided that we embed its definition within the envisaged umbral restyling.

We note indeed that the use of this operator allows the following alternative definition:
\begin{equation*}
	{_{L}D_{x}} = \frac{d}{d X}, \quad \text{where} \quad X = x\, c_{z},
\end{equation*}
as a consequence of the fact that
\begin{equation*}
	\frac{d X^{n}}{d X} = n\, X^{n-1}.
\end{equation*}
If we now use the following special notation
\begin{equation*}
	\exp\left[{\lambda\, {_{L}D_{x}}}\right] = \exp\Big[\frac{\lambda}{c_{z}}\frac{d}{d x}\Big],
\end{equation*}
by keeping separate the actions of the derivative and umbral operators, we find, e. g.,
\begin{align*}
	\exp\left[{\lambda\,{_{L}D_{x}}}\right]\, X^n &= \exp\Big[{\frac{\lambda}{c_{z}} \frac{d}{d x}}\Big]\, c^n_{z} \, x^n \frac{1}{\Gamma(1+z)}\Big\vert_{z=0} \\ 
	&= c_{z}^{n} \left(x + {\lambda\,c_{z}^{-1}}\right)^n  \frac{1}{\Gamma(1+z)}\Big\vert_{z=0} \\
	& = (\lambda + x\, c_{z})^n  \frac{1}{\Gamma(1+z)}\Big\vert_{z=0}\\ 
	&= L_{n}( \lambda,x) = \lambda^n L_{n}\left(-{x/\lambda}\right) = \exp\left[{\lambda \,D_{x}^{(0)}}\right]\, \frac{x^n}{n!},
\end{align*}
or
\begin{align*}
	\exp\left[\lambda\, {_{L}D_{x}} \right] \E^{a\, X} & = \exp\Big[\frac{\lambda}{c_{z}} \frac{d}{d x}\Big] \,\exp\left[a \,x\, c_{z}\right]\,\frac{1}{\Gamma(1+z)}\Big\vert_{z=0}\\  
	&= \exp\left[a \,c_{z}\left(x + {\lambda\,c_{z}^{-1}}\right)\right] \,\frac{1}{\Gamma(1+z)}\Big\vert_{z=0} \\ 
	& = \E^{ a\,\lambda}\, \exp[a\, x\, c_{z}]\, \frac{1}{\Gamma(1+z)}\Big\vert_{z=0} = \E^{a \,\lambda}\, I_{0}(2\sqrt{a\,x}).
\end{align*}
Observe also that 
\begin{align*}
	\exp\left[\lambda\, {_{L}D_{x}} \right]\, e^{a \,X} & = \exp\left[\lambda\, {_{L}D_{x}} \right]\sum_{r=0}^{\infty} \frac{(a \,X)^r}{r!}\\  
	&= \sum_{r=0}^{\infty} \frac{a^r}{r!} \exp\left[{\lambda\, {_{L}D_{x}} }\right]\, X^r \\ 
	&= \sum_{r=0}^{\infty} \frac{a^r}{r!} \exp\left[\lambda \,D_{x}^{(0)}\right]\, \frac{x^r}{r!}  \\
	& = \exp\left[{\lambda\, D_{x}^{(0)}}\right] \sum_{r=0}^{\infty} \frac{(a\,x)^r}{(r!)^2}  = \exp\left[{\lambda\, D_{x}^{(0)}} \right]\, I_{0}(2\sqrt{ax}).
\end{align*}

The same definition can be used in an even more raffishly way by noting that 
\begin{align*}
	\exp\left[{\lambda\, {_{L}D_{x}}} \right]\,\E^{-x} &= \exp\Big[{\frac{\lambda}{c_{z}} \frac{d}{d x}} \Big]\,\frac{1}{1 + x c_{z}} \frac{1}{\Gamma(1+z)}\Big\vert_{z=0} \\
	&= \frac{1}{1 + c_{z}\left(x + {\lambda}\,{c_{z}^{-1}}\right)} \,\frac{1}{\Gamma(1+z)}\Big\vert_{z=0} \\
	& = \frac{1}{1 + \lambda}\, \frac{1}{1 + (x \,c_{z})/(1 + \lambda)} \,\frac{1}{\Gamma(1+z)}\Big\vert_{z=0} \\
	&= \frac{1}{1 + \lambda} \exp\left({-\frac{x}{1 + \lambda}}\right) = \exp\left[{\lambda\, D_{x}^{(0)}} \right]\E^{-x}.
\end{align*}

\end{document}